\documentstyle[emulateapj]{article}
\def\etal{{\sl~et\,al.~}}

\begin{document} 

\title {Using the Comoving Maximum of the Galaxy Power Spectrum to Measure 
Cosmological Curvature}

\author
{Tom Broadhurst, Andrew H.~Jaffe}
\affil{Department of Astronomy, University of California, Berkeley, CA  94720}


\begin{abstract}
  
  The large-scale maximum at $k\simeq0.05\;h/{\rm Mpc}$ identified in
  the power-spectrum of galaxy fluctuations provides a co-moving scale
  for measuring cosmological curvature. In shallow 3D surveys the peak
  is broad, but appears to be well resolved in 1D, at $\sim
  130\;h^{-1}{\rm Mpc}$ ($k=0.048\;h/{\rm Mpc}$), comprising evenly
  spaced peaks and troughs. Surprisingly similar behaviour is evident
  at $z\sim3$ in the distribution of Lyman-break galaxies, for which
  we find a $5\sigma$ excess of pairs separated by $\Delta
  z\simeq0.22\pm0.02$, equivalent to $85\;h^{-1}{\rm Mpc}$ for
  $\Omega=1$, increasing to $170\;h^{-1}{\rm Mpc}$ for $\Omega=0$,
  with a number density contrast of 30\% averaged over 5 independent
  fields.  The combination, $3.2\Omega_m-\Omega_{\Lambda}\simeq0.7$,
  matches the local scale of 130$\;h^{-1}{\rm Mpc}$, i.e. 
  $\Omega=0.2\pm0.1$ or $\Omega_{m}^{flat}=0.4\pm0.1$ for the
  matter-dominated and flat models respectively, with an uncertainty
  given by the width of the excess correlation. The consistency here
  of the flat model with SNIa based claims is encouraging, but
  overshadowed by the high degree of coherence observed in 1D compared
  with conventional Gaussian models of structure formation. The
  appearance of this scale at high redshift and its local prominence
  in the distribution of Abell clusters lends support to claims that
  the high-$z$ `spikes' represent young clusters.  Finally we show
  that a spike in the primordial power spectrum of
  $\delta\rho/\rho\simeq0.3$ at $k=0.05\;h/{\rm Mpc}$ has little
  effect on the CMB, except to exaggerate the first Doppler peak in
  flat matter-dominated models, consistent with recent observations.

\end{abstract}


\section
{Introduction}

    Methods for measuring cosmological parameters are more readily
conceived than realized in practice. Recently however, it has become
clear that both SNIa lightcurves and CMB temperature fluctuations
provide reasonably model-independent and complimentary means of
constraining cosmological curvature (Goobar \& Perlmutter 1995,
Eisenstein, Hu \& Tegmark 1998).  The angular scale of the first CMB
Doppler peak limits combinations of $\Omega$ and $\Lambda$. At smaller
angles more geometric information may be extracted but is subject to
greater physical uncertainty and complex foregrounds. The utility of
distant Type Ia supernovae has been convincingly demonstrated
(Perlmutter\etal 1998,1999), but concerns remain.  Local supernovae
are largely discovered in luminous dusty late-type galaxies,
deliberately targeted for convenience.  In the distant field,
supernovae are weighted towards lower luminosity galaxies, sometimes
undetected because of the volume-limited nature of the selection.
Hence it is not clear that both samples should share the same family
of light curves. The physical origin of the $\sim$15\% luminosity
residual remaining after correction is not understood, nor indeed the
light-curve correlation itself. Furthermore, some evolution is natural
via progenitor metalicity, although the redshift
independence of both the residual variance and the distribution of
event durations is reassuring.  An upward revision of the distant SNIa
luminosities of only 15\% would render a cosmological constant
unnecessary (Perlmutter\etal 1999).

 Here we attempt a simple model-independent measurement of curvature,
using the spatial scale of the maximum in the power spectrum of galaxy
perturbations. A broad maximum is naturally expected below
$k\sim0.1\;{\rm Mpc}/h$, corresponding to the Horizon at the epoch of
matter-radiation equality, up to which high frequency power is
suppressed relative to an initially rising power-law spectrum, bending
the spectrum over at higher frequency.  Observations have established
a maximum at $k\sim 0.05\;h/{\rm Mpc}$ ($\sim130\;h^{-1}{\rm Mpc}$)
from wide angle galaxy and cluster surveys (Bough \& Efstathiou
1993,1994, Einasto\etal 1997, Tadros, Efstathiou \& Dalton 1998,
Guzzo\etal 1998, Einasto\etal 1998, Hoyle \etal 1998, Scheuker \etal
1999). In particular a pronounced peak is evident in local cluster
samples (Einsato\etal 1998) and in the deprojection of the large APM
survey (Bough \& Efstathiou 1994). In all these estimates the high
frequency decline is steeper than expected (Feldman, Kaiser \& Peacock
1995) and together with the relatively small amplitude fluctuations
implied by CMB at very low k (Smoot \etal 1992) requires the existence of a
maximum between 0.01$<$k$<$0.1.

  The low frequency peak in the projected 3D power-spectrum and its
coincidence in scale with the excess large power detected in lower
dimensional redshift surveys (Broadhurst\etal 1990,1992,1999; Landy\etal
1995, Small, Sargent \& Hamilton 1997, Tucker, Lin, Schectman 1998) 
encourages the possibility that the test proposed here can be explored
already, by simply comparing pencil beam surveys at high and low
redshift. Power on these large scales comoves with the expansion,
providing a means of measuring curvature. For example, at $z=0$ the
redshift interval corresponding to a scale of $130\;h^{-1}{\rm Mpc}$
is $\Delta z=0.043$, but this stretches by a factor of 4--8 in redshift to
$\Delta z\simeq 0.25$ at $z=3$, increasing with $\Omega$ and $\Lambda$.

  Here we describe the evidence for a preferred large scale of $\Delta
z\simeq0.22$ in the fields of high redshift galaxies presented by
Steidel (1998) and in the Hubble Deep Field (Adelberger\etal 1998). We
examine this in light of the increasing local evidence for a maximum
in the power spectrum at 130$\;h^{-1}{\rm Mpc}$. We obtain the locus
of $\Omega$ and $\Lambda$, under the assumption that the preferred
scale in these datasets is the same. Finally we discuss the
implication of a spike in the primordial power spectrum for the CMB.

\section {Low Redshift Structure}
  
  Several independent wide angle surveys of galaxies and clusters
allow the projected 3D power spectrum, $P_3(k)$ to be estimated on
scales large enough to cover the predicted low frequency
roll-over. These surveys all show a sharp decline in power at
$k>0.1\;{\rm Mpc}/h$, and if deep enough, evidence of either a peak,
or a maximum at $k\sim0.05\;{\rm Mpc}/h$ (Baugh \& Efstathiou 1993,
Einasto \etal 1995, 1998, Tadros \& Efstathiou 1996, Hoyle \etal
1999, Schuecker \etal 1999). This despite the relatively sparse sampling
and fairly shallow depth, so that the peak may be better resolved in
larger ongoing surveys, where the binning in $k$ can be made finer,
and the isotropy of Fourier amplitudes examined around $k=0.05\;{\rm
Mpc}/h$ for any coherent or generally non-Gaussian behaviour.

 Consistent with this, redshift surveys directed at the Bo{\"o}tes
void and towards the Phoenix region and the Shapley supercluster also
show a pronounced pattern of large wall/voids separated by
$\sim130\;h^{-1}{\rm Mpc}$ (Kirshner\etal 1981, 1987, 1990,
Quintana\etal 1988, Small, Hamilton \& Sargent 1997). Large strip
surveys in two dimensions also contain excess power around $\sim
100\;h^{-1}{\rm Mpc}$ (Landy\etal 1995, Vettolani\etal 1995).  This is
despite the narrowness of the strips, of only a few Mpc in width,
which must therefore underestimate the real radius of voids (Tucker,
Lin \& Shectman 1999).

 The transverse extent of the peaks and troughs identified at the
galactic poles by Broadhurst \etal (1990) are now known to span much
more than the width of the original beams, which were of order only
the correlation length ($5\;h^{-1}{\rm Mpc}$). The closest northern peak
at v=7000km/s intersects the transverse `great wall' structure near
the Coma cluster, as revealed in the maps of De-Lapparent, Geller \&
Huchra (1986), which extends over $100\;h^{-1}{\rm Mpc}$.  Wider
angle redshift surveys at the galactic poles (of diameter
$100\;h^{-1}{\rm Mpc}$) have clearly confirmed and strengthened the
early finding of a scale of $130\;h^{-1}{\rm Mpc}$ spanning the
redshift range $z<0.3$ (Broadhurst \etal 1999) revealing a sinusoidal
alternating pattern of peaks and troughs spanning the redshift range
$z<0.3$ and hence a correspondingly narrow concentration of power in
one dimension $P_1(k)$, at $k=0.045\;h/{\rm Mpc}$. Independent support
for the reality of this pattern is found in the coincidence of these
peaks with the local distribution of rich clusters (Broadhurst 1990,
Bahcall 1991, Guzzo\etal 1992) and in a southern field close to the
pole (Ettori, Guzzo \& Tarengi 1997).
  
  In comparing the 1D distribution with the projected 3D power spectrum
it is important to keep in mind that the wide angle redshift surveys
do not have sufficient volume and sampling density to construct the
real 3D power spectrum $P_3(\vec{k})$ on 100$\;h^{-1}{\rm Mpc}$
scales, so interpretations are based on $P_3(|k|)$, which is the mean
power averaged over solid angle. Hence, it is only sensible to compare
the amplitude of $P_3(|k|)$ with $P_1(k)$ if the power is known to be
isotropic with $\vec{k}$. A non-Gaussian distribution leads to
``hotter'' and ``colder'' spots at a given frequency, which may
average out in projection but generate a larger variance in 1D pencil
beams. This is particularly true of course for highly coherent
structure (e.g. Voronoi foam, Icke \& Van De Weygaert 1991) , where the
3D variance can be sub-Poissonian on scales larger than the coherence
length.
 
\section{High Redshift Structure}
 
  The most puzzling aspect of the distant dropout galaxies is the
appearance of sharp peaks in the redshift distribution 
(Adelberger\etal 1998, Steidel 1998), resembling the situation at low
redshift, implying at face value little growth of structure.  A
conventional interpretation of these peaks resorts to ``bias''
(Wechsler \etal 1998) which has come to represent a flexible
translation between the observed structure and the relatively smooth
mass distribution of standard simulations, so that the observed peaks
are interpreted as rare events destined to become massive clusters by
today (Wechsler\etal 1998).  High biases have been claimed for the
Lyman-break population on the basis of the amplitude of small scale
clustering at $z\sim3$ (Adelberger\etal 1997,
Giavalisco\etal 1998). The occurrence of such peaks is enhanced with a
steep spectrum by a reduction of high frequency `noise' (Wechsler et
al 1998) but this gain is offset if the steepness is attributed to low
$\Omega$, since then a given redshift bin corresponds to a larger
volume, and hence a greater proper density contrast (Wechsler et al
1998). We may regard the existence of regular spikes at low and high
redshift as evidence for a revision in our understanding of large
scale structure, indicating perhaps that initial density fluctuations
are not Gaussian distributed such as may be implied by the observed
lack of evolution of the number density of X-ray selected clusters
(Rosati \etal 1998). The baryon isocurvature model (Peebles 1997) more
naturally accommodates both the early formation and the frequent
non-Gaussian occurrence of high density regions (Peebles 1998a,b).

 We analyze the fields of Lyman-Break galaxies of Steidel (1998) and
Adelberger(1998). These include 4 fields of $\sim 100-200$ galaxies
and a smaller sample of redshifts in the Hubble Deep Field
direction. The fields are $\sim10$Mpc in width by $\sim 400$Mpc in the
redshift direction and include over 600 redshifts in the range
$2.5<z<3.5$ histogrammed in bins of $\Delta z=0.04$.  In Figure~1 we
plot the pair counts and correlation function, assuming that galaxies
are evenly distributed within the narrow redshift bins. A clear excess
is apparent on a large scale, corresponding to a preference for
separations of $\Delta z=0.22$. A pair excess is also seen at twice
this separation (Fig~1) indicating phase coherence along
the redshift direction.  This behaviour is an obvious consequence of
the regular peaked structure visible in the redshift histograms of
four of the five fields (Fig~1). In a bin of 10Mpc, the number of pairs at the
peak is $\sim 1220$ compared with an expected $\sim 1060$, 
representing a $4.6\sigma$ departure from random (Fig~1).

The observed redshift interval may be related to comoving scale 
by the usual formula for a universe with negligible radiation
energy-density (Peebles 1993),
\begin{eqnarray}\label{eq:deltaw}
  \Delta w &=& 3000\; h^{-1} \;{\rm Mpc}\; \int_z^{z+\Delta z} {dz\over E(z)} 
  \nonumber\\
  &\simeq& 3000\; h^{-1} \;{\rm Mpc}\; {\Delta z\over E(z)}
\end{eqnarray}
with
\begin{equation}
  \label{eq:Ez}
  E(z) = \sqrt{
    \Omega (1+z)^3 + \Omega_\Lambda + (1-\Omega-\Omega_\Lambda)(1+z)^2}.
\end{equation}
Whereas standard candles at moderate redshift (SNIa) measure
$\sim(\Omega - \Omega_\Lambda)$, and CMB anisotropies measure
$\sim(\Omega + \Omega_\Lambda)$, these observations considered here
measure $\sim E(3)$ or $\sim (3\Omega-\Omega_\Lambda)$ which lies
between these two locii, thus adding complimentary information.

 With this, the $\Delta z=0.22$ scale of the peak corresponds to
$85\;h^{-1}{\rm Mpc}$ for $\Omega=1$, doubling to $170\;h^{-1}{\rm
Mpc}$ for $\Omega=0$. Flat cosmologies with a positive $\Lambda$ fall
between these limits. If we constrain $\Delta w = 130\; h^{-1}{\rm
Mpc}$ we find $48\Omega-15\Omega_\Lambda \simeq 10.5$. This gives
$\Omega=0.2$ for an open universe ($\Omega_\Lambda\equiv0$) or
$\Omega=0.4$ for a flat universe ($\Omega+\Omega_\Lambda\equiv1$),
with an uncertainty of only $0.1$ in $\Omega_m$, given by the 15\%
width of the excess correlation (Figure~1). This scale, if borne out
by subsequent redshift data, is certainly consistent with the flat
model preferred by SNIa data with $\Omega_m^{flat}=0.3$ (Perlmutter \etal
1998,1999).

\section {Excess Power and the CMB}

 One can imagine two broad classes of physical mechanisms that might
be responsible for the excess power required to fit this data. The power might be a truly primordial feature in the power
spectrum. In inflationary scenarios, for example, this could be
generated by the proliferation of super-horizon bubbles in a suitably
conspiratorial inflaton potential (La 1991, Occhinero \& Amendola
1994). On the other hand, excess power could be due to causal
microphysics in the universe after the 130~$\;h^{-1}{\rm Mpc}$ scales
enter the horizon. In other words, the power could be added by the
transfer function. A high baryon density naturally imparts large scale
``bumps and wiggles'' to the power spectrum (Peebles 1998a,b,
Eisenstein\etal 1997, Meiksin\etal 1998) in particular if matter
density fluctuations are created at the expense of radiation (Peebles
1998a,b). An even more radical possibility is that the universe is
topologically compact, and that we are seeing evenly-spaced copies of
a small universe with an extent of only $\sim130\;h^{-1}{\rm Mpc}$
although this scale seems too small to accommodate the unique and 
relatively distant (z=0.18) cluster A1689 (Gott 1980).

In light of this it is interesting to explore the implications of any
excess power on CMB temperature fluctuations.  Eisenstein\etal (1997)
has examined in what way conventional adiabatic models maybe stretched
to match the power spectrum of Broadhurst\etal (1990), demonstrating
that such models do not naturally account for a scale of
130$\;h^{-1}{\rm Mpc}$ in the mass distribution.

Our approach is to simply add power in the primordial spectrum at a
fixed wave number, using a modified version of CMBFAST code (Seljak \&
Zaldarriaga 1999), to simulate temperature fluctuation spectra. A
primordial spike of power will effect the CMB directly through the
projection of three-dimensional features.  A narrow band of power is
added, $\Delta^2(k) = k^3 P(k)/(2\pi^2)$. The amplitude is set so that
$\int d\ln{k}\; \Delta^2(k) \simeq 0.1$, the value of the correlation
function at the peak and harmonics equivalent to at 30\% density
contrast. We ignore bias, which may conceivably be large,
lowering the peak amplitude; we also ignore the possibility of
non-Gaussian fluctuations which would certainly diminish the
angle-averaged power.

If the value of the power spectrum at the $k=0.05\; h^{-1}{\rm Mpc}$
peak is already large, as with COBE-normalized $\Lambda$ models, an
additional peak of the above strength has negligible effect.  However,
COBE-normalized models with $\Omega=1$ and $\Lambda=0$ have
comparatively low power at this scale, hence, the effect on the final
$C_\ell$ and $P(k)$ is significant. Of course, the precise width and
location of the peak, not yet at all well-constrained by the CMB data,
also affects the relative power in the peak versus the underlying
``smooth power-law'' spectrum.  This is illustrated in Figure~2. The
spike is seen to raise the amplitude of the first Doppler peak by
nearly a factor of two for the sCDM model.

 Note, although it is premature to interpret the claims of the various
CMB experiments for and excess of power at $C_\ell=200$ (Netterfield
et al 1997, Tegmark 1998) without proper treatment of the covariance
matrix of errors and foreground subtraction, the indications of a
large amplitude for this peak are not inconsistent with the degree of
boosting (Fig.~2).  Others, Gawiser and Silk 1998, have noted that the
whole aggregate of current $C_\ell$ and $P(k)$ data is extremely
well-fit---much better than the fit to standard models---by an
adiabatic inflationary power spectrum with sharp bump like that
considered here.

\section {Conclusions}

 Two interesting results emerge from the above comparison of structure
in the local and distant pencil beam data. Firstly both samples of
galaxies show regular large scale power confined to a narrow range of
frequency. Secondly, by matching these scales we obtain values of
$\Omega$ and $\Lambda$ in good agreement with the SNIa claims.  These
findings may of course be regarded as remarkable coincidences, unlikely
to occur by chance in a clumpy galaxy distribution (Kaiser \& Peacock
1991).

Distinguishing between the open and flat solutions found here requires
pencil beam data at a third redshift. Optimally this redshift turns
out to be convenient for observation, $z\sim0.7$, where the ratio of
$\Delta z$ between the open $\Omega=0.2$ and $\Omega_m^{flat}=0.4$
cases is maximal, differing by 13\%, and may be explored with
existing data.

 Since locally the excess power at 130$\;h^{-1}{\rm Mpc}$ is most
prominent in cluster selected samples, we may conclude that the peaks
in the high-$z$ sample correspond to proto-clusters at $z\sim 3$. This
conclusion is independent of the high-bias interpretation which also
regards the spikes as young rich clusters.

Understanding the physics behind these spikes is complicated by a
conspiracy of scales. We have seen in Fig.~2 that the
$k\simeq0.05\;{\rm Mpc}/h$ spike produces a feature
in the CMB power spectrum at nearly exactly the position of the first
doppler peak. In the simple model presented here---a primordial spike at 
this position---this is merely a coincidence, however unlikely.
That these scales are so closely matched seems yet more unlikely in the
light of another fact, that the {\em required} peak in the matter power
spectrum is at nearly the position of the {\em expected} peak due to the
passage from radiation to matter-domination.

If the acoustic peaks in the CMB power spectrum are instead a
signature of physics at a somewhat later epoch: the sound horizon at
recombination (set by $c/\sqrt{3}$) then that these scales are
nearly coincident (within an order of magnitude or so of 100
$\;h^{-1}{\rm Mpc}$) is a consequence of the particular values of the
cosmological and physical parameters.  Thus, if we change the initial
conditions to increase the amplitude of the matter power spectrum near
the equality scale, we also increase the CMB temperature power
spectrum at roughly the scale of the first acoustic peak.  That is, in
neither case do we add an unexpected peak, but merely increase the
amplitude and ``sharpness'' of an expected one.

Of course, the inexplicable coincidence is that the feature in the
power spectrum appears close to the expected matter-radiation equality
peak, but with an amplitude much too large to be explained by standard
theories. In addition, we must also note that merely adding a peak to
the mean power spectrum cannot account for the nearly periodic
structure observed in 1D. A real theoretical explanation must account
for the complicated non-Gaussian properties of this distribution.

 The excess 1D power identified by Broadhurst\etal (1990) will soon be
subject to easy dismissal with large 3D redshift surveys. However,
whether or not it transpires that universal large scale coherence
exists, the test proposed here is still viable in principle using the
general predicted turnover of the 3D power spectrum, requiring larger
volume redshift surveys.  In particular comparisons of local and
distant clusters over a range of redshift will be particularly useful
given their sharply peaked and high amplitude power spectrum..

\acknowledgments

We thank Richard Ellis, Alex Szalay, Gigi Guzzo, Alvio Renzini, Eric
Gawiser, Saul Perlmutter and Martin White for useful conversations.

\fontsize{10}{14pt}\selectfont

\begin{figure}[t] 
\epsscale{1}
\plotone{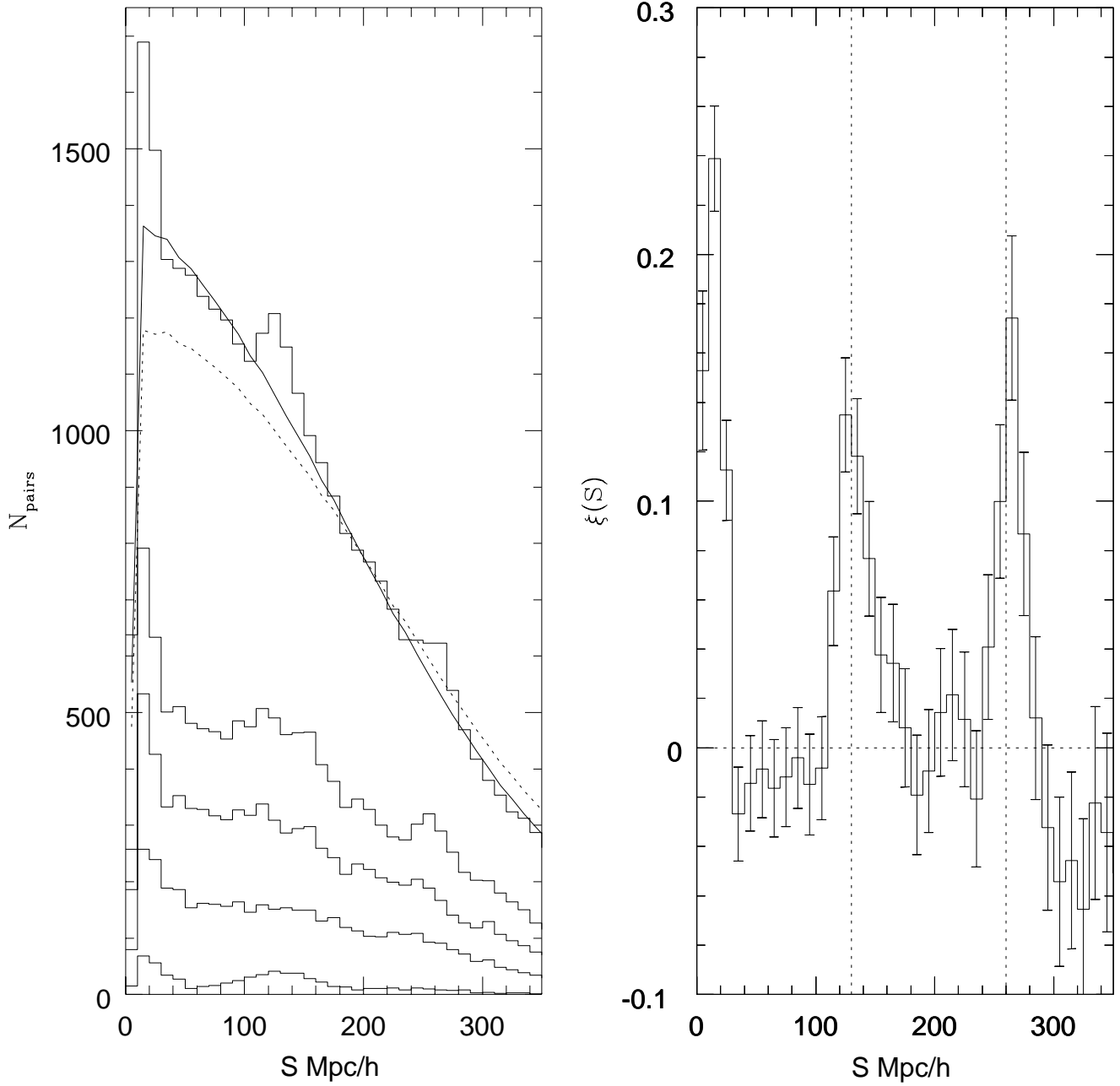}
\caption{The co-added pair counts of 5 fields of galaxies at $z\sim3$
  compared with the random expectation of a modified selection function
  based on (solid curve) Adelberger\etal 1998 (dotted curve). The contributions of the five fields are also histogrammed, ordered by number of redshifts 
 for clarity. The right
  hand panel shows the pair weighted correlation function with 
  obvious excesses at 130$\;h^{-1}{\rm Mpc}$ and at
 double this scale, 260$\;h^{-1}{\rm Mpc}$, ($\Omega=0.2$) implying 
 large scale coherence.}
\end{figure}
\begin{figure}[t] 
\epsscale{1}
\plotone{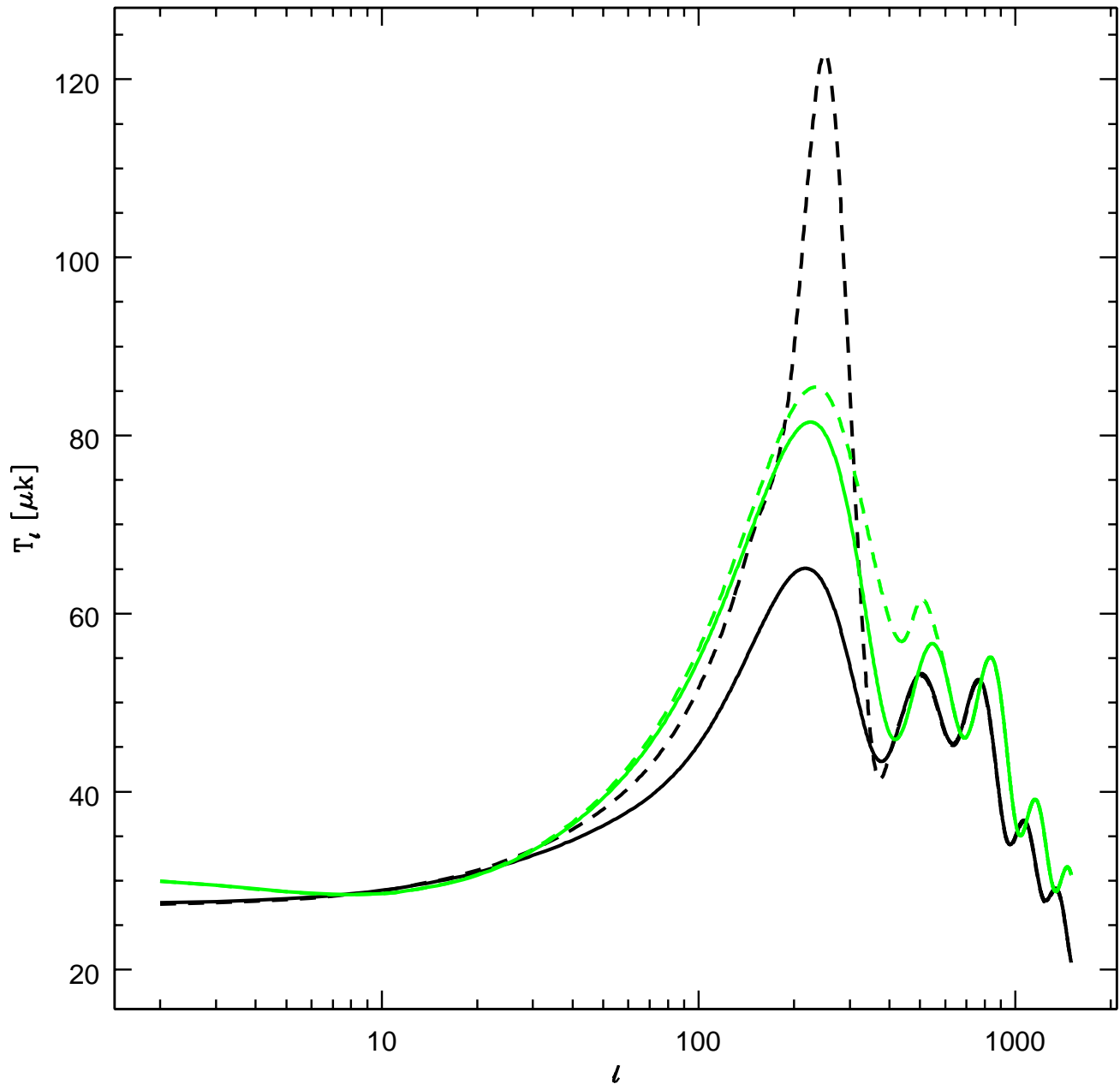}
\caption{Models for the CMB temperature fluctuation
  spectrum including flat (with no cosmological constant) standard CDM
  models (green) and a $\Lambda$ dominated model (black). Standard $n=1$
  primordial power spectra are shown with solid lines and the addition
  of a peak at $k=0.05\; h/{\rm Mpc}$ with dashed lines.  The highest
  dashed and the lowest solid curves are the SCDM model; the middle pair
  or the $\Lambda$ model. Note that the peak generates a relatively
  large enhancement of the first doppler peak in the sCDM case.}
\end{figure}
\end{document}